\documentclass[aps,preprint,superscriptaddress,showpacs]{revtex4}
\usepackage{amsmath,bm}
\usepackage{graphicx}

\begin{document}

\title{Electron orbital valves made of multiply connected armchair carbon
nanotubes with mirror-reflection symmetry: tight-binding study} 
\author{Gunn Kim}
\affiliation{School of Physics and Astronomy, FPRD, and Center for Theoretical Physics, Seoul National University, Seoul 151-747, Korea}
\affiliation{Department of Physics, North Carolina State University, Raleigh, NC 27695, USA}
\author{Sang Bong Lee}
\author{Hoonkyung Lee}
\author{Jisoon Ihm}
\email[corresponding author. E-mail:\ ]{jihm@snu.ac.kr}
\affiliation{School of Physics and Astronomy, FPRD, and Center for Theoretical Physics, Seoul National University, Seoul 151-747, Korea}
\date{\today }

\begin{abstract}

Using the tight-binding method and the Landauer-B\"{u}ttiker conductance formalism,
we demonstrate that a multiply connected armchair carbon nanotube with a mirror-reflection symmetry can sustain 
an electron current of the $\pi$-bonding orbital
while suppress that of the $\pi$-antibonding orbital
over a certain energy range.  
Accordingly, the system behaves like an electron orbital valve 
and may be used as a scanning tunneling microscope to probe pairing symmetry in $d$-wave superconductors 
or even orbital ordering in solids which is believed to occur in some transition-metal oxides.

\end{abstract}
\maketitle

\section{introduction}
An electronic state in atoms, molecules, or solids has a particular spatial character usually represented by
an orbital or a linear combination of such orbitals well-established in quantum chemistry. 
Since an enormous number of different kinds of orbitals coexist and they usually overlap in a real material, 
it is difficult to select or probe a particular orbital in real solids.
Recently, as fabrication techniques of nanometer-sized material units such as carbon nanotubes (CNTs) are developed,
control over individual electronic states in a material is improving quite remarkably.\cite{McEuen,Tans,Martel,Javey,Fuhrer}
Here we show theoretically that, in a multiply connected armchair carbon nanotube (MCACN) system with a mirror-reflection symmetry,
we can produce an electron current of one particular orbital character (the $\pi$-bonding $p$ orbital or simply $\pi$ orbital)
and suppress the current flow of the other type of electrons (the $\pi$-antibonding $p$ orbital or simply $\pi^{*}$ orbital)
over a significantly wide range of energy.
It means that the system is metallic for one kind of orbitals and insulating for another orbitals,
in close analogy to a half-metal which is metallic for, say, up-spins and insulating for down-spins.
The electron tunneling probability from this system to a molecule 
critically depends on the orbital character of the molecule
because of a different degree of overlap between the wave functions of this system and the molecule.
The system behaves like an electronic orbital valve (filter) which selectively transmits the $\pi$ orbitals only.
Furthermore, it may be used as a scanning tunneling microscope (STM) to probe the orbital characters of molecules
or pairing symmetry in $d$-wave superconductors
or even orbital ordering in solids which is believed to occur in some transition-metal oxides.\cite{YTokura}

In this paper, we report interesting transport properties of an MCACN structure as shown in Fig. 1,
where a single metallic tube is branched off into two smaller arms and then they merge into one.\cite{Grimm,GKim}
Our model comprises two leads of semi-infinite metallic (6,6) CNTs 
and two arms of finite (3,3) tubes in between.
Figures 1(a) and 1(b) show MCACNs with two arms of finite (3,3) tubes of the same length and
of two different arm lengths, respectively.
In fact, there is an unpublished report\cite{MTerrones} that such a ``needle's eye configuration" has been formed,
though not intentionally fabricated, by electron irradiation on nanotubes
and observed in the Transmission Electron Microscopy image of nanotubes. 
Six heptagons are contained in each of two junction regions 
where a thicker tube and two thinner tubes are joined. 
These heptagons near each junction area act as a scattering center
for electrons incident from the lead [the (6,6) nanotube].
The length of the arm region is represented by the number 
($L$) of periodic units of the armchair CNT as shown in Fig. 1. 
In the following study, we require that the structure possesses the mirror-reflection symmetry 
with respect to a plane containing the axes of the (3,3) and the (6,6) tubes (i.e., the plane of Fig. 1 
containing the tube axis).

\section{computational details}
An energy band consisting of $\pi$ orbitals and another band of $\pi^{*}$ orbitals 
are crossing at the Fermi energy ($E_F$) in the armchair CNT
and they are usually admixed when defects are introduced 
into the perfect nanotubes breaking symmetries of the structure. 
However, the situation is different if a mirror-reflection symmetry 
with respect to the plane containing the tube axes is maintained.
The $\pi$ orbital is even with respect to the mirror-reflection 
whereas the $\pi^*$ orbital is odd with respect to it.
Then $\pi$ and $\pi^*$ transport channels remain unmixed and mutually independent,
which is a rigorous result of quantum mechanics.
The electronic structure is described by the single $\pi$-electron tight-binding
Hamiltonian
$$H = V_{pp\pi} \sum_{<i,j>}(a^{\dagger}_{i}a_{j}+H.c.),$$
where $<i,j>$ denote the nearest neighbor pairs,
the hopping integral $V_{pp\pi}=-2.66$ eV~\cite{Blase} and
the on-site energy is set to zero.
In our scattering-state approach, we solve the Schr\"odinger equation
for the whole system by matching the solutions of the tight-binding equation
in the arm with that of each lead (the perfect nanotube) region at the interface on the left and right.
For a given incoming electron wave from the left lead as an initial condition, 
we obtain the transmitted electron emerging on the right lead (as well as the reflected electron traveling to the left). 
Conductance is obtained from the Landauer-B\"{u}ttiker formula,
$G(E)=G_{0} {\rm Tr}({\bf t^{\dagger}t})$, where $G_0$ is the conductance quantum 
($=2e^2/h$) and ${\bf t}$ is the transmission matrix.\cite{Lan,Butt}

\section{results and discussion}
The conductance as a function of the incident electron energy is displayed in Fig. 2. 
Two arm lengths are assumed to be the same in Fig. 2(a).
The total conductance is decomposed into two nonmixing contributions of 
$\pi$ and $\pi^*$ channels, $G = G_0 (T_{\pi} + T_{\pi^*})$, where
$T_{\pi}$ and $T_{\pi^*}$ are the transmission probabilities of the two.
Since the $\pi$ orbitals do not have phase variation in the circumferential direction around the tube
irrespective of $n$ in ($n$,$n$) tubes,
$\pi$ orbitals in the lead parts [(6,6) tubes] have 
almost perfect match with those in the arms [(3,3) tubes],
and the transmission is close to unity in a wide energy range.
On the other hand, the sign of $\pi^*$ orbitals alternates rapidly over the perimeter of the tube.
The $\pi^*$ orbitals in the lead part cannot match with those of two smaller arms simultaneously.
Therefore, the transmission of $\pi^*$ orbitals through the system is mostly suppressed 
except for the near resonant energy levels as displayed in Fig. 2(a).

To understand the positions and the linewidths of these conductance peaks,
we investigate the structure of the wave functions in more detail.
Besides the linear $\pi$ and $\pi^*$ bands crossing at $E_F$, 
there exist a conduction band at 1.3 eV and above, and a valence band at $-$1.3 eV and below in the (6,6) tube.
Because of the presence of the heptagonal carbon rings at the junction,
donor-like states tend to be produced according to well-known H\"uckel's ($4n+2$) rule for stability.\cite{Solomons}
The localized state due to topological defects (7-membered rings) at Y-junctions is most pronounced 
if the energy level is close to (and below) the aforementioned conduction band minimum (CBM) at 1.3 eV.
The sharp peaks closest to the CBM in Fig. 2(a) corresponds to this state and has a narrow lineshape
owing to the relatively well-localized (long lifetime) character of such an origin.
The resonant states further away from the CBM are less localized and thus broader.
They may be regarded as states confined in the arm region broadened 
by the interaction with the continuum of metallic states in the lead regions on both sides.
We also note that the interference between a narrow and a broad level gives rise to asymmetric resonant peak structure
which is known as ``Fano Resonance."\cite{Fano}
We previously found the same features in a larger MCACN which comprises two leads of semi-infinite metallic (10,10) CNTs
and two arms of finite (5,5) tubes of the same length in between,\cite{GKim}
but the phenomenon of the orbital valve was not emphasized at the time,
which will be studied in detail below.

We now examine the orbital valve (filter) in the conductance of the system.
In all cases we study, there exists a finite energy range 
where the conductance of $\pi^*$ electrons is almost zero
while that of $\pi$ electrons is not reduced appreciably.
For instance, the transmitted electrons of $\pi^*$ orbital for the MCACN with two different arm lengths
of $L_1=11$ and $L_2=12$ in Fig. 1(b)
constitute less than 5\% of the total transmitted electrons in the energy range
between $-$0.2 and 0.1 eV as shown Fig. 2(b).
Figure 3 shows charge density plots of scattering states in the MCACNs 
and explains two origins causing almost perfect reflection of $\pi^*$ electrons in the wide energy range.
As depicted in Figs. 3(a) and 3(b), the incident wave at $E=0.77$ eV from the left lead is reflected 
at the left Y-junction, regardless of lengths of two arms. 
In contrast, the incident wave at $E=0$ eV from the left lead is reflected at the right Y-junction
because of the destructive interference by two paths of different lengths 
as seen in Fig. 3(c), in analogy to Young's double-slit experiment.
Here $E=0.77$ eV is chosen because the MCACN with the same arm length has the transmission zero at this energy,
and $E=0$ eV is chosen because a realistic orbital filter could operate without gate bias 
for electrons near the Fermi energy.

This phenomenon, which we call ``orbital filtering,"
is a counterpart of spin filtering in half-metals such as La$_{1-x}$Sr$_{x}$MnO$_3$.
Half-metal has been extensively studied for possible applications as a spin valve in spintronics.
The difference is that, while the density of states (DOS) of up-spins is finite 
and that of down-spins is zero at the Fermi level in half-metals,
the electrical conductance (rather than the DOS) of $\pi$ orbitals is finite 
and that of $\pi^*$ orbitals is zero in the MCACN
though the DOSs of both $\pi$ and $\pi^*$ orbitals are nonzero at the Fermi level.
Such a non-equilibrium distribution of current-carrying states (imbalance between $\pi$ and $\pi^*$ states)
may be maintained over the relaxation length (at least several nm) of the electron phase beyond the arm region.
The MCACN acts as an orbital filter and this concept can be used 
for a further fundamental research on the electronic character 
in nanostructures or an application to an orbital switch or orbital sensor using nanotubes.
We note that there is increasing interest in identifying orbital degrees of freedom
in various transition-metal oxides such as La$_{1-x}$Sr$_x$MnO$_4$ and Ca$_{2-x}$Sr$_x$MnO$_4$.\cite{YTokura}

People have used, for instance, cross-polarized optical microscopes to probe orbital ordering
with the micron-scale resolution.
Using the MCACN system as an STM, in principle, it is possible to probe electrically the orbital character
of a molecule or orbital ordering in solids.
The principle is identical to the spin valve system in the spin switch or spin sensor.
If an orbital in a sample matches and overlaps well with the $\pi$-orbital of the probe (nanotube),
the tunneling current increases. If the orbital character in the sample differs 
(or points to an unfavorable direction) from the probe orbital, the current decreases or vanishes
because the tunneling current is proportional to the absolute square of the overlap integral. 
For most materials, occupation of a specific orbital at the expense of vacating other orbitals is an extreme
non-equilibrium situation and the relaxation to equilibrium distribution
(equal occupation of the same energy states) occurs typically in a time scale of less than 1 fs.
This is a severe restriction on the practical experiment for the orbital probe.
A great advantage of using carbon nanotubes for the single-orbital (i.e., $\pi$ orbital only) current source is that
the electron phase coherence length (or phase relaxation length) $L_{\phi}$ 
in carbon nanotubes is known to be unprecedentedly long
(several tens of nm even at room temperature).\cite{Schonenberger,Stojetz}
Having a source of constant phase, we can probe orbital character of a molecule 
or orbital ordering on the surface of solids.
The quantitative analysis of the tunneling data requires a more detailed numerical integration
because the $\pi$-orbital state actually consists of atomic $p$ orbitals which have sign change
in the radial direction of the nanotube (though not in the circumferential direction).

Figure 4 shows a pedagogical example of the square of the overlap integral 
between the $\pi$ state of the carbon nanotube
and an atomic {\it d} orbital ($\left| \langle \Psi_d | \pi \rangle \right|^2$).
For definiteness, we choose a single $3d_{3z^2-r^2}$ and $3d_{x^2-y^2}$ orbitals of a hydrogen-like atom 
(the atomic number $Z=10$; the effective core charge\cite{Clementi1,Clementi2} 
for {\it d}-electrons in a Mn atom).
The {\it xy} coordinates of the figure indicate the position of the center of the nanotube with respect to the atom.
The distance in the {\it z}-direction between the edge of the tube and the atom is chosen to be 7 \AA.
The axis of the tube is perpendicular to the {\it xy}-plane.
This simple example demonstrates that the current of $\pi$ orbitals from the nanotube 
(acting as an STM tip) can unambiguously
distinguish between $3d_{3z^2-r^2}$ and $3d_{x^2-y^2}$ orbitals possibly even at room temperature.
Due to the orbital sign change in the circumferential direction of the nanotube, in contrast, 
the square of the overlap integral between the $\pi^*$ state of the carbon nanotube
and an atomic {\it d} orbital ($\left| \langle \Psi_d | \pi^* \rangle \right|^2$) does not represent
the shape of the {\it d} orbital.

\section{conclusion}
We have carried out tight-binding calculations for conductance
of MCACNs with the mirror symmetry.
It is found that the electron current of one particular orbital character ($\pi$ orbital) is sustained while 
the current flow of other electrons ($\pi^{*}$ orbital) is suppressed  over a wide range of energy.
It implies that the system is metallic for one kind of orbitals and insulating for another orbitals, in close
resemblance to a half-metal used in spintronics.
Since the electron tunneling probability from this system 
to a molecule or s solid critically depends on the orbital characters of the molecule or surface atoms, 
the system may be used as a probing tip of the orbital character of molecules or solids.

\begin{acknowledgments}
This work was supported by the Samsung SDI-SNU Display Innovation Program,
the KRF Grant No. KRF-2005-070-C00041, the MOST through NSTP Grant No.M1-0213-04-0001,
and the SRC Program (Center for Nanotubes and Nanostructured Composites) of MOST/KOSEF.
Computations are performed through the support of the KISTI.
\end{acknowledgments}

\newpage
\begin{center}
\LARGE{[Figure Captions]}

\end{center}

FIG. 1. Ball-and-stick models of MCACNs with two arms of finite (3,3) tubes (a) of the same arm length ($L=12$) and
(b) of two different arm lengths ($L_1=11$ and $L_2=12$), respectively.
Two leads consisting of (6,6) tubes are attached on the left and right.

FIG. 2. Conductance plots as a function of the incident energy {\it E}~~for the MCACNs
(a) of the same arm length ($L=12$) and (b) of two different arm lengths ($L_1=11$ and $L_2=12$), respectively, 
in units of conductance quantum ($2e^2/h$).

FIG. 3. Charge density plots of scattering states for MCACNs
of the same arm length ($L=12$) in (a) and of two different arm lengths ($L_1=11$ and $L_2=12$) in (b) and (c), respectively.
The energies of the incident waves are 0.77 eV for both (a) and (b), and 0 eV in (c),
exhibiting different electronic distribution of (c) compared with (a) and (b).
The darker sphere indicates higher charge density.

FIG. 4. The pattern of the overlap integral between the (6,6) carbon nanotube $\pi$ orbital 
and an atomic orbital, $\left| \langle \Psi_d | \pi \rangle \right|^2$.
Here $\Psi_d$ is chosen as (a) a single $3d_{3z^2-r^2}$ orbital or (b) a single $3d_{x^2-y^2}$ orbital of 
a hydrogen-like atom ($Z=10$).
The {\it xy} coordinates are the position of the center of the nanotube relative to the hydrogen-like atom in \AA.
The brightness of the figure is the intensity in arbitrary units.

\newpage
\begin{figure}[t]
  \centering
  \includegraphics[width=11cm]{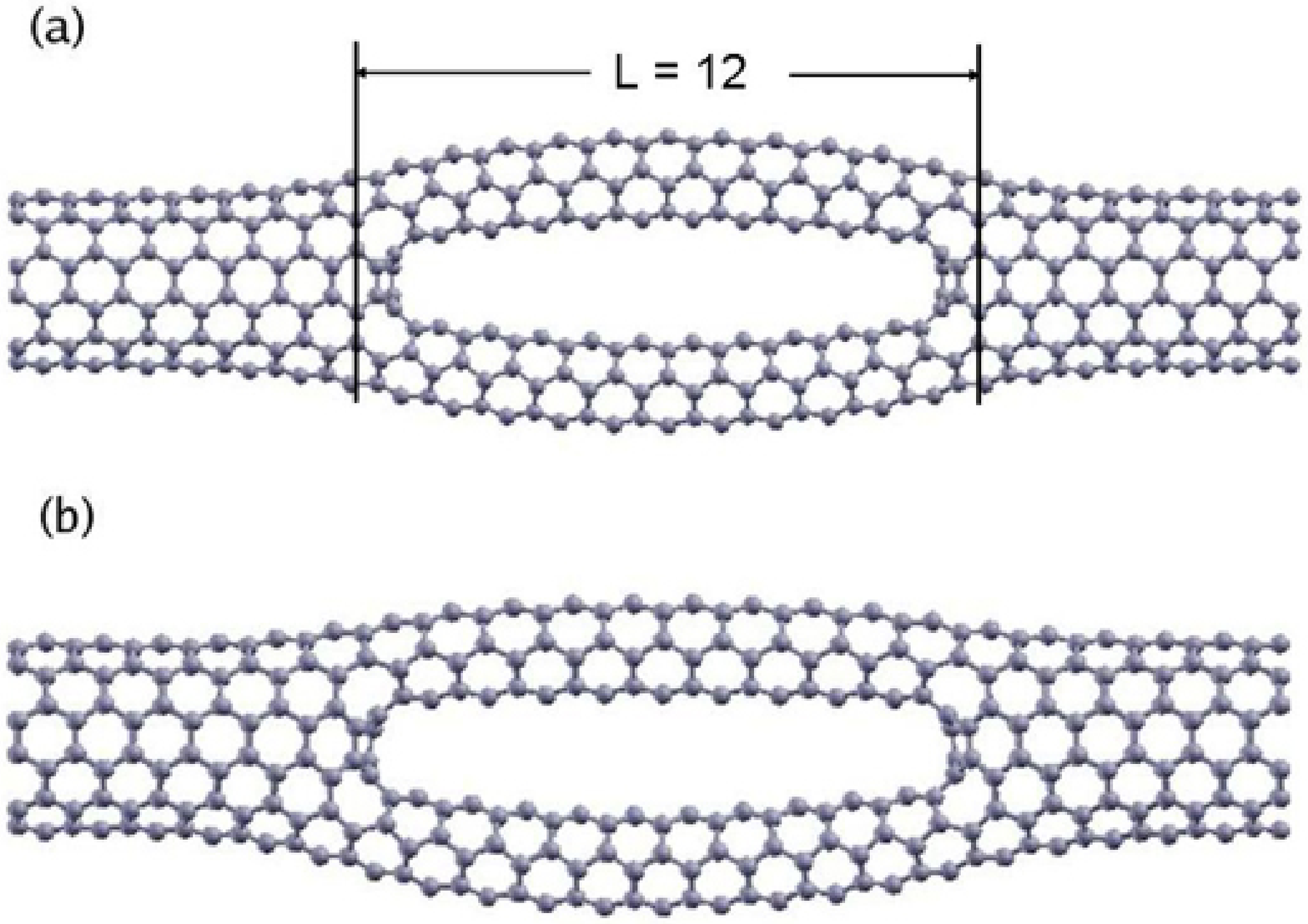}
  \label{model}
\end{figure}
\begin{center}
\LARGE{Figure 1}

\LARGE{G. Kim, S.B. Lee, H. Lee and J. Ihm}
\end{center}

\newpage
\begin{figure}[t]
  \centering
  \includegraphics[width=11cm]{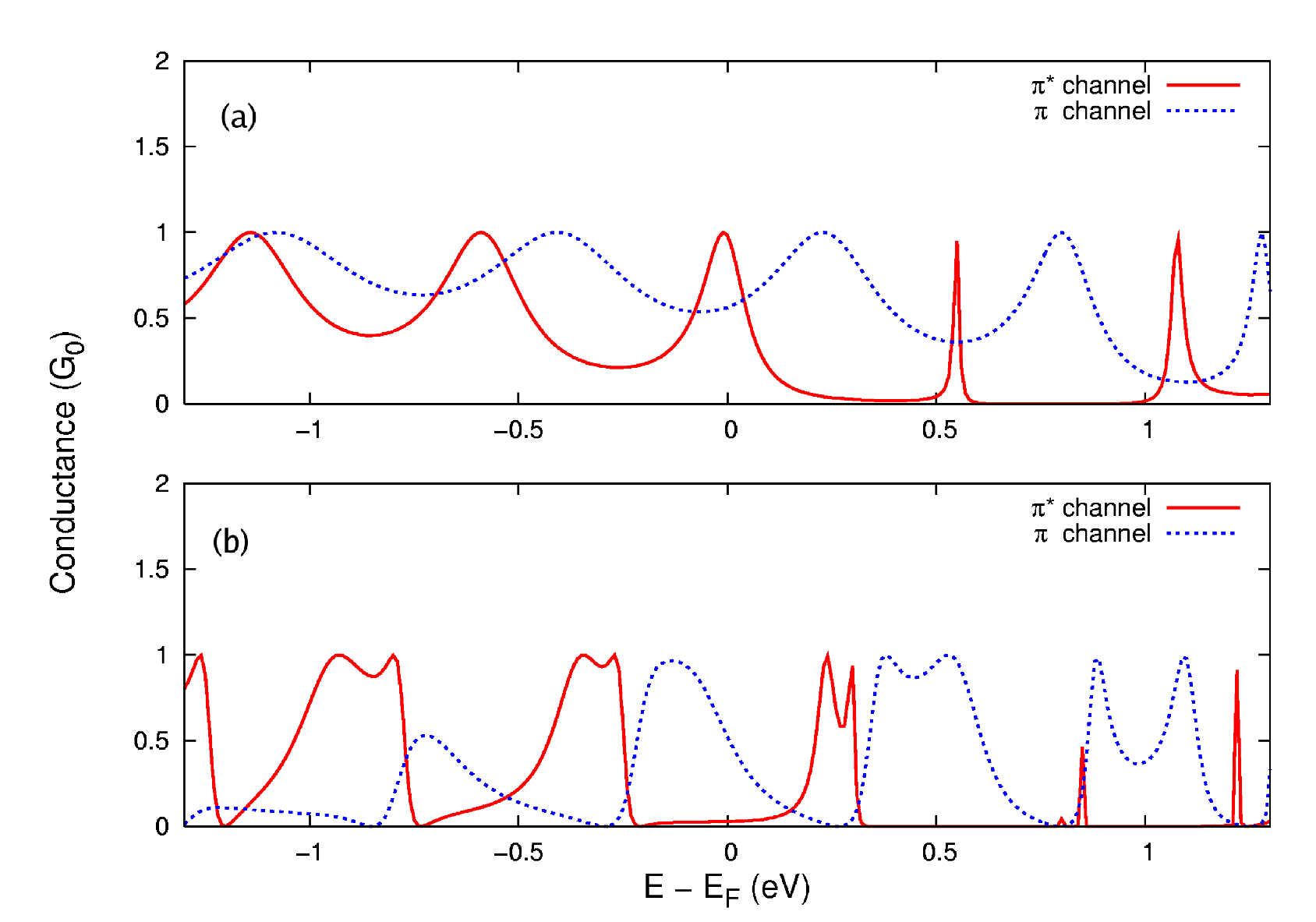}
  \label{conductance}
\end{figure}
\begin{center}
\LARGE{Figure 2}

\LARGE{G. Kim, S.B. Lee, H. Lee and J. Ihm}
\end{center}

\newpage
\begin{figure}[t]
  \centering
  \includegraphics[width=11cm]{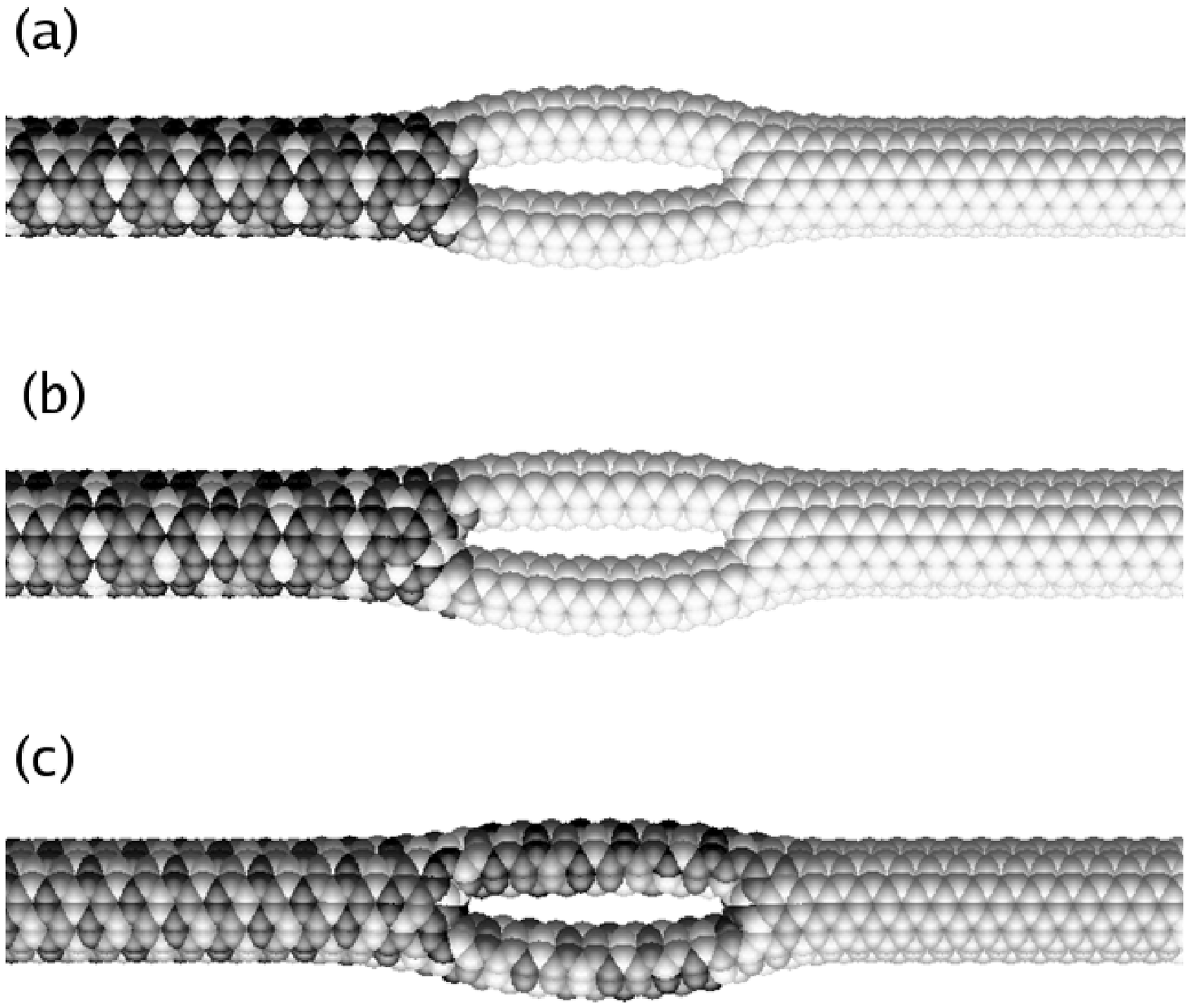}
  \label{wavefunction1}
\end{figure}
\begin{center}
\LARGE{Figure 3}

\LARGE{G. Kim, S.B. Lee, H. Lee and J. Ihm}
\end{center}



\end{document}